\documentstyle[prl,aps,epsf]{revtex}

\begin{document}
\draft
\preprint{}
\title{Evolution of photoemission spectral functions in doped transition 
metal oxides}
\author{H. Kajueter and G. Kotliar}
\address{
Serin Physics Laboratory, Rutgers University, Piscataway, NJ 08855 USA}
\author{D.D. Sarma and S.R. Barman}
\address{Solid State and Structural Chemistry Unit, Indian Institute of
Science, Bangalore 560 012, INDIA}
\date{\today}
\maketitle
\begin{abstract}
We discuss the  experimental photoemission and inverse photoemission
of early transition metal oxides, in the light of the dynamical mean
field theory of correlated electrons which becomes exact
in the limit of infinite dimensions.
We argue that a comprehensive description of the experimental data
requires spatial inhomogeneities and present a calculation
of the evolution of the spectral function in an inhomogenous  system
with various degrees of inhomogeneity.
We also point out that comparaison of experimental results and
large d calculations require that the degree of correlation and
disorder is larger  in the surface than in the bulk.
\end{abstract}
\narrowtext
\twocolumn

There has been a resurgence of interest in doped transition metal oxides
following the discovery of high T$_c$ oxides \cite{BM}. 
Early transition metal oxides, containing Ti and V, have attracted
particular attention in very recent times \cite{Tokura,Fuji,inoue,Fuji1}.
La$_x$Sr$_{1-x}$TiO$_3$ \cite{Tokura} is such an 
example where the carrier concentration can be varied in a
controlled fashion.  
Varying $x$ tunes the electron configuration
continuously between 0 and 1, with LaTiO$_3$ ($x$ = 0) being a 
Mott-Hubbard insulator with a 3$d^1$ configuration and SrTiO$_3$ ($x$ = 1)
being a band insulator with a 3$d^0$ configuration.
For large values of $x$, the transport
and magnetic properties \cite{Tokura} as well as the experimentally
observed spectral weights behave as a correlated doped Mott
insulator.  For small values of $x$ ($\approx$ 0), the system
represents small amount of electron doping in a $3d^0$ configuration,
and thus, it should behave as a {\it doped band insulator}. Indeed
transport and magnetic properties of such
compounds \cite{Tokura,transport}  exhibit  essentially
free-electron like behavior.

On the thoeretical side, it  has been shown
that the Hubbard model in the limit of large dimensions offers
a semiquantitative description of the evolution of the
transport and thermodynamic
data  as a function of carrier concentration, in these systems.
The consistent modelling of the {\it photoemission spectroscopy,}
however, is  more problematic and is the subject of this paper.

We first  summarize the main experimental facts that a theory 
should describe. 
1) In a pioneering paper Fujimori {\it et al.} \cite{Fuji}
investigated the spectral
function of systems with a $d^1$ configuration.
They  observed that there are two features in the UPS spectra,
a low energy feature near the Fermi edge, and an intermediate energy
feature  around 1.4 eV which we refer to as the coherent and the incoherent
parts of the spectral function respectively.
The metal-to-insulator transition at filling $n=1$ is
driven by the transfer of spectral weight from the low energy feature
to the incoherent part of the spectra. 
Inoue and collaborators confirmed this picture in a more careful
study of the CaSrVO$_3$ system \cite{inoue}. 
This behavior can be understood in terms of the solution
of the infinite dimensional Hubbard model at half filling\cite{rozenberg},
 but the parameters
required  to fit the data  produce specific heat and susceptibility
enhancements which are inconsistent with the thermodynamic data in these
systems. \cite{inoue2}

2) In a subsequent publication Fujimori and collaborators addressed \cite{Fuji1}
the doping dependence of the coherent and incoherent features
in La$_x$Sr$_{1-x}$TiO$_3$ with various values of  $x$ .
They found that as one dopes the Mott insulator, the weight 
in the incoherent and in the  coherent features grow.
Furthermore, they observed that
the  incoherent feature persists up to fairly large values
of doping, when one would have expected nearly free electron
behavior.
Similar behavior was reported in the $Y_{1-x} Ca_{x} Ti O_3$ 
system.
The behavior at large doping (small electron number)
has  been confirmed in careful studies
of the  SrTiO$_{3-\delta}$  system. \cite{tsukuba}

In contrast to the experiments described in
the previous paragrpah, no theoretical model ever came close
to explaining these observations.
In particular  the large d Hubbard model
which successfully accounted   the  transport and thermodynamic data in
the doped Mott insulator, completely fails in reproducing 
the experimentally  observed evolution of
the spectral function as one varies the carrier
concentration.

3) Sarma and collaborators revisited the limit of small density of electrons.
They performed photoemission and inverse photoemission spectroscopy
on SrTiO$_{3- \delta}$ \cite{Sarma}, where the oxygen deficiency $\delta$ 
introduces electrons into the stoichiometric $d^0$ system.
The photoemission part of the spectra confirmed the findings of Fujimori 
{\it et al.} with an incoherent feature around 1.4 eV and a coherent feature
at low frequencies, while the inverse photoemission spectra were in 
complete agreement with band structure calculations.

We  stress again that these results 
are very surprising, since in a dilute system the interactions
are always renormalized to small values so both photoemission
and inverse photoemission should be predicted correctly
by the band structure calculations.

4) The discrepancy between the parameters that are necessary
to fit the bulk thermodynamical and transport properties on one hand and the
highly surface sensitive 
photoemission spectroscopy on the other hand, which was alluded to earlier, 
suggests that the behavior near the surface may be different
than in the bulk. While this issue is technically difficult to be probed on the
dilute $d^1$ systems such as La$_x$Sr$_{1-x}$TiO$_3$ or SrTiO$_{3- \delta}$, 
it can be effectively addressed
in the denser systems such as the La$_x$Ca$_{1-x}$VO$_3$ system with electron
configuration varying between 1 and 2.
Sarma and coworkers \cite{Maiti} performed a systematic study of the evolution
of the spectral function as a function of photon energy and
observed that the incoherent features get weaker
while the coherent features become stronger as the 
probing depth of the spectroscopic technique increases by varying the photon
energy suitably.
 
Taking into account all these disparate experimental observations, we believe
that to obtain a coherent picture of the experimental situation
in these systems one has to recognize, that 1) a minimal model of 
the spectroscopic
data   in these systems has to recognize that the surface experiences
a different degree of correlation and disorder than the bulk.
2) The ultra-violet photoemission spectroscopy   can only be modelled 
successfully by models that include {\it spatial }
inhomogeneities.
We argue that the existent data are good enough to rule out
homogeneous models but  cannot determine the length scale over
which  the inhomogeneity takes place.

In the framework of the infinite dimensional Hubbard model
we present  calculations of the evolution of the
spectral function  with increasing doping in
an interacting disordered system and argue that they are in
good agreement with all the   experimental observations
cited above.
Whereas substituted compounds
have been extensively investigated in recent times, the role of
disorder has almost always been neglected; thus our findings have
important implications for a large number of materials.
The present work also brings into focus the 
different degrees of disorder and correlations at
the surface and in the bulk; while this is indeed to be expected and should
be taken into account when interpreting experimental data, but has not received any attention before.

In order to motivate our model we note that the
Hubbard model represents a homogeneous system, whereas in real
systems electron (or hole) doping is brought about by chemical
substitution; {\it this invariably introduces disorder or
inhomogeneities at some length scale}. 
There are several reasons why the strengths of the correlations
and the disorder are different in the bulk than at the surface.
Atoms at the surface experience a different coordination number
than in the bulk, therefore the ratio of the interaction strength
to the average hopping matrix element is larger at the surface
than in the bulk. The strength of the disorder is related
to the spatial fluctuations in the effective potential
experienced by an atom.  In the limit of infinite coordination
off-diagonal disorder becomes a non fluctuating quantity , 
i.~e.~it is self averaging. So the effective strength of off-diagonal
disorder {\it increases} with decreasing
effective dimensionality.

 In this paper  we propose that
it is essential to use 
 a spatially  {\it inhomogeneous} picture
to model  the surface of these  systems.
 The basic idea is that in these systems the
electrons experience different local environments.  The angle
integrated  photoemission spectrum, as a local probe, captures a
suitable average of the Greens function over the local environments.
In principle the degree of inhomogeneity can vary between two limits.
If the system is disordered on a microscopic scale
 (termed microscopic inhomogeneity here),
but homogeneous
on a macroscopic scale, i.e. the impurities do not form clusters, a
model of uncorrelated disorder is appropriate. 
If there is a
tendency of the defects to form clusters, this may reinforce the
natural tendency of strongly correlated systems to phase separate as
pointed out in ref\cite{Emery} and
a  model of macroscopic inhomogeneities is more suitable. 
In this paper we use
the  dynamical mean field theory that becomes exact
in the limit of infinite spatial
coordination \cite{rev}  to make these ideas more
quantitative. 
We study the evolution of photoemission spectra in
spatially inhomogenous systems which we model using
{\it diagonal disorder}.
Off diagonal disorder which would correspond more
closely to the physical situation  disappears in the limit
of large dimensions, unless one introduces very special
distributions of hopping matrix elements \cite{dobrosavlevic}.
We find that the photoemission spectrum as a function of $x$ to be very
different from the predictions of the Hubbard model and very similar
to what is observed experimentally.

We study  a  Hubbard
model with diagonal disorder:  
\begin{displaymath}
H = - \sum_{(ij)} t_{ij\sigma} C^{+}_{i\sigma}C_{j\sigma} + \sum_{i}
(\epsilon_{i} - \mu) C^{+}_{i\sigma}C_{i\sigma} + \sum_{i}
n_{i\uparrow} n_{i\downarrow} U
\end{displaymath}
To get a bounded density of states and a well defined limit as the
lattice coordination gets large, we take the hopping matrix elements
$t_{ij}$ to be defined on a Bethe lattice and scale them as
$\frac{t}{\sqrt{d}}$.  
The full
bandwidth $2 D = 4 t$, is   estimated to be 2.5 eV  to fit
our band 
structure calculations.
The $\epsilon_{i}$ are random variables which take  the values
$\epsilon_{A}$ with probability $P_{A}$ and $\epsilon_{B}$ with
probability $P_{B}$; $P_A$ and $P_B$ are given by the composition, as
$2 \delta$ and $(1- 2 \delta)$ in SrTiO$_{3-\delta}$ and $x$ and
$(1-x)$ in La$_x$Sr$_{1-x}$TiO$_3$ respectively.  
Supercell calculations for SrTiO$_{2.875}$ and SrTiO$_{2.75}$
have shown
that several bands are pulled down compared to SrTiO$_3$ due to the
presence of vacancies by about 1.5 - 3.0 eV which provides an
estimate for $(\epsilon_B - \epsilon_A)$. We use $(\epsilon_B -
\epsilon_A)$ $\sim 2.4$ eV; however, the qualitative features of the
calculations do not depend on this exact value. $U_A = U_B = $ 4.6 eV
was estimated from earlier spectroscopic analysis.  The chemical
potential $\mu$ is chosen so that the number of $d$ electrons is
equal to $x$ in La$_x$Sr$_{1-x}$TiO$_3$ or $2\delta$ in
SrTiO$_{3-\delta}$.
The   diagonal disorder introduces randomness that  
survives  the limit of large dimensionality, and
captures  the basic physics that different sites
experience different  local  enviroments.

The mean field theory which is exact  in the  $d \rightarrow \infty$
limit determines the local  Greens functions at the A and B sites as
the local $f$ electron Greens function of an  Anderson impurity model
\begin{eqnarray}
H & = & \sum_{k} \epsilon_{k} a^{+}_{k\sigma} a_{k\sigma} +
U f^{+}_{\uparrow} f_{\uparrow}
f^{+}_{\downarrow} f_{\downarrow}\nonumber \\
 & & +  \sum_{k} V_{k}(a^{+}_{k\sigma}f_{\sigma} + f^{+}_{\sigma}
a_{ko})
 +  \epsilon_{f} f^{+}_{\sigma} f_{\sigma}
\label{anderson}
\end{eqnarray}
Here the $f$ level position  $\epsilon_{f}$ is given by  $\epsilon_{A}-\mu$ and
$\epsilon_{B}-\mu$, respectively.  The hybridization function
$\Delta(i \omega_n)$
\begin{equation}
 \Delta(i \omega_n)= 
\sum_{k}
\frac{V^{2}_{k}}{(i\omega_{n} - \epsilon_{k})}
\label{hyb}
\end{equation}
has to be determined self consistently from the local Greens
functions via the equation:
\begin{displaymath}
\nonumber
\Delta(i\omega_n) = t^2\ G_{av}(i \omega_n)
=t^2\ ( P_{A}\ G_{A}(i\omega_n) + P_{B}\ G_{B}(i\omega_n))
\end{displaymath}

It is to be noticed that the A and B sites are connected by the same
hybridization function which depends on the local physics at both
sites. Thus we have a coupled system of equations.  To solve the mean
field equations we use the recent extension of the IPT
method\cite{IPT} to extract real frequency information.

Figure \ref{fige4} (a) shows the total as well as the partial spectral
functions for small filling ($n=0.2$\/).
Since $\epsilon_B>>\epsilon_A$, the particle concentration on the $B$
sites is very small. Thus the corresponding spectral function is
only slightly renormalized by interactions. It
basically consists of
a semicircle around $\epsilon_B-\mu$ with a small
correlation induced peak at $\epsilon_B-\mu+U$.
On the $A$ sites, in contrast, we find almost one particle per lattice
site. Therefore $\mbox{Im} G_A(\omega)$ exhibits features similar to
a slightly doped Mott insulator, i.~e.~there are two Hubbard bands
around $\epsilon_A-\mu$ and $\epsilon_A-\mu+U$ plus a resonance peak
at zero frequency. Notice, that the lower Hubbard band of
$G_A(\omega)$ produces a correlation induced feature in the
averaged Greens function $G_{av}(\omega)$, as observed in the experiments.
This maximum grows with increasing doping as can be seen from the
evolution of the spectral function shown in the left column of 
figure \ref{fige6}. In the limit $n\to 1$, there are only $A$ sites
left and the incoherent 
feature observed at negative frequencies becomes the
lower Hubbard band of a pure Mott insulator.

We now turn to the case where the inhomogeneities
occur on a macroscopic scale.
In this
scenario the density  of electrons is inhomogeneous on a more
macroscopic scale.  A fraction $f_1$ is in an insulating phase with
the  local concentration of the Mott insulating phase  $n_{ins}=1$
and a fraction $1-f_1$  with the metallic concentration $n_{met}$.
The total density $n={f_1}n_{ins}+(1- {f_1})n_{met}$, and the
measured Greens function is given by
\begin{equation}
\label{separate}
\bar{G} = {f_1} G_{n_{ins}} + (1-{f_1}) G_{n_{met}}
\end{equation}
$G_{n_{ins}}$ and $G_{n_{met}}$ can again be calculated  using the
mean field theory now in the respective homogeneous phases.  To
explore this scenario we took $n_{met}=.1$ (which corresponds to a
chemical potential $\mu=-.78$) and varied the density $n$ by varying
the volume fraction ${f_1}$ in equation (\ref{separate}).
The total and the partial spectra for this case are shown in
figure \ref{fige4} (b). Except for the absence of a resonance peak
at zero frequency the picture is similar to the case of 
microscopic disorder discussed above. In particular, the lower Hubbard
band of the insulating phase produces an incoherent feature at
negative frequencies, which grows with increasing filling
(see right column of figure \ref{fige6}).
The experimentally measured quantitiy is a convolution
of  spectral distribution at negative
(photoemission) or positive (Bremsstrahlung isochromat) frequencies
with a resolution  function.
\ref{fige5}
 (a) and (b)  shows  this  convolution   
 that can now  be compared directly 
with the experimental data of Sarma {\it et al.} \cite{Sarma}.
The data are in excellent 
agreement with the experimental spectra. 
For example, the calculation correctly
predicts the existence of an intense incoherent feature at about 1.3
eV arising from the local Greens function for site A with a band-like
coherent feature near $E_F$ due to the local Greens function for site
B. This is in perfect agreement with the experimentally observed
features in the photoemission spectra of SrTiO$_{3-\delta}$ 
\cite{Sarma} 
and La$_x$Sr$_{1-x}$TiO$_3$ \cite{Fuji1}.  Moreover, it is clear from
the calculated results that unoccupied parts of the spectral function
will be entirely dominated by coherent band-like features arising
from the site B as observed in the experiment \cite{Sarma}.

On the other hand, we would like to stress that the experimentally
measured  specific heat,
a bulk property, cannot be described using the parameters obtained
from the (surface sensitive) photoemission experiments. 
Experimental measurements \cite{Tokura}
show that $\gamma=C/T$ is enhanced as the
Mott insulator is approached with increasing filling.
It is clear that this observation cannot be accounted for by phase
separation, because the metallic phase is assumed to be a dilute
system hardly showing any correlation effects and the insulator does not
contribute to the linear coefficient of the specific heat.
( we assume that the insulating state gives a  negligible   contribution
to the specific heat )
The scenario of microscopic disorder in contrast vastly overestimates
this quantity. This can be understood by the fact that the electrons
get trapped on the $A$ sites so that the local particle density
$n_{A}$ is close to one and much larger than the average particle
density $n$ (see above).
As has been established in reference \cite{Gabi1},
the specific heat can, however, be described by the 
{\em homogeneous} Hubbard model.

To summarize,  in this paper we  discuss a model for the
high-energy spectroscopic data of compounds with close
to 3$d^0$ configuration.
We presented explicit calculations of the spectral function of
correlated electrons at different dopings in an inhomogeneous picture.
We considered two cases; in one the inhomogeneity is present only on
a microscopic scale while in the other the system was assumed to be
macroscopically inhomogeneous. Both models give similar angle integrated
spectra which therefore 
cannot  determine the length scale
over which the system is inhomogeneous; however, the calculations are in
excellent agreement with known experimental results for such and similar
systems.  
The  spectra of the inhomogeneous models
are qualitatively different from that of
a homogeneous system, suggesting the need to go beyond
the standard theories to describe spectral functions in a class of doped
transition metal compounds.

FIGURES:
\begin{itemize}
\item
Fig. 1
{Spectral functions of the disordered
Hubbard model for density n=.2, U=3.7 in the case of
(a) microscopically and (b) macroscopically
inhomogeneous  disorder. The energy scale is given in
units of $D$ (= 1.25 eV). }\label{fige4}
\item
Fig. 2 
{Evolution of the spectral functions of the disordered
Hubbard model for $U=3.7$,
\mbox{$\epsilon_b-\epsilon_a=1.9$,} and increasing
filling $n$; left column: homogeneous disorder, right column:
inhomogeneous disorder} \label{fige5}
\item
Fig. 3
{Theoretical spectra of the disordered
Hubbard model for density $n=.2$, $U=3.7$, and $\epsilon_b-\epsilon_a=1.9$
  in the case of
(a) microscopic and (b) macroscopic
inhomogeneities.
The dashed and dotted curves indicate the partial contributions from
the A and B sites.
The occupied parts of the spectral
functions are expanded for clarity, as shown in the figures. }\label{fige6}

\end{itemize}

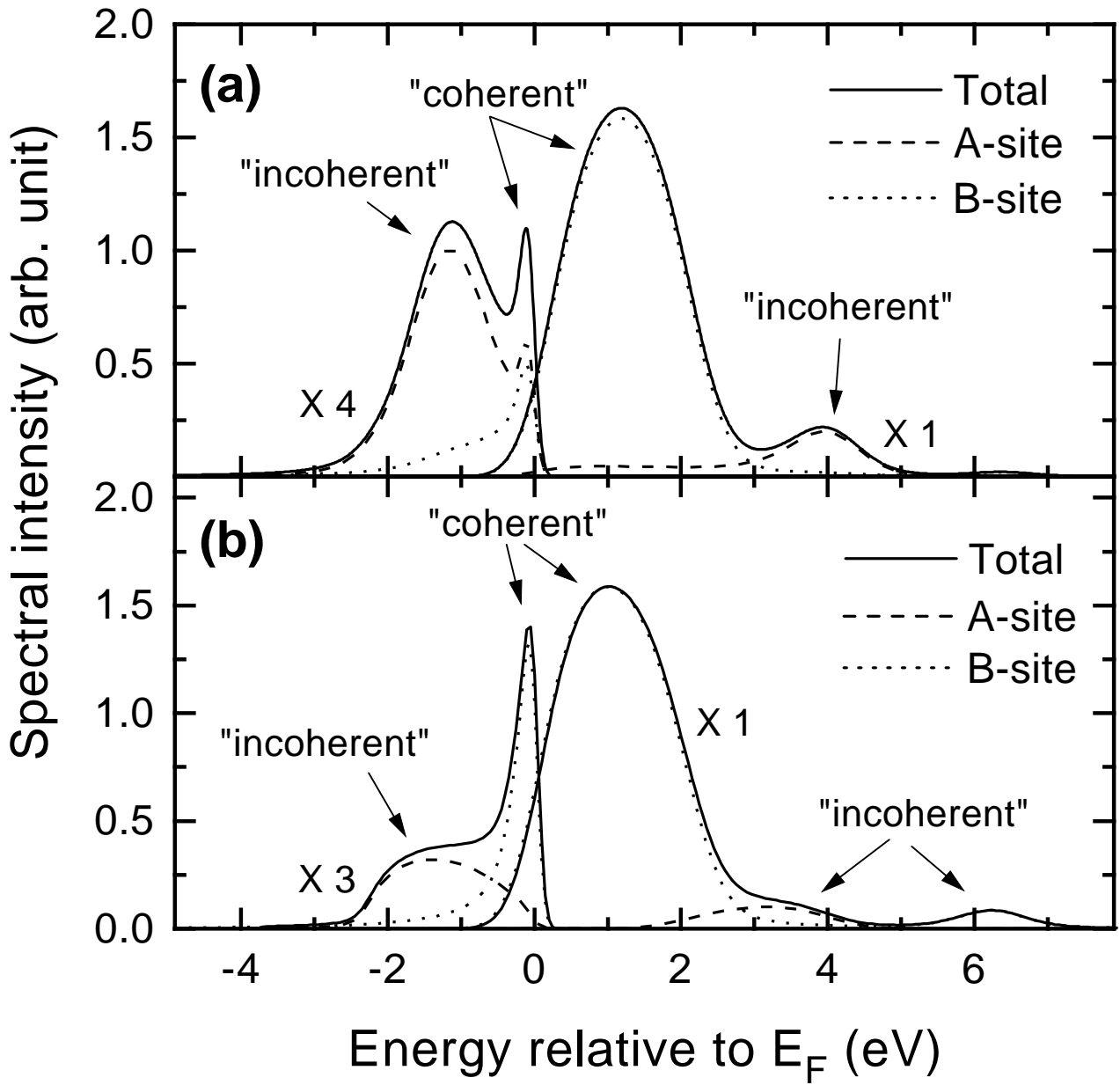
\end{document}